\documentstyle[12pt,epsfig]{article}

\makeatletter
\def\slash#1{{\mathpalette\c@ncel{#1}}} % TeXbook, bottom of p360
\makeatother

\newcommand\beq{\begin{eqnarray}}
\newcommand\eeq{\end{eqnarray}}

\newcommand\la{\langle}
\newcommand\ra{\rangle}

\def\ellslash{\rlap/{\mkern-1mu \ell}}

\def\shat{\hat{s}}
\def\that{\hat{t}}
\def\uhat{\hat{u}}

\begin{document}
\vspace*{7mm}
\begin{center}
{\Large \bf 
Hyperon Polarization from Unpolarized $pp$ and $ep$ Collisions}

\vspace{0.5cm}
 {\sc Yuji~Koike}
\\[0.3cm]
%\vspace*{0.1cm}
{\it Department of Physics, Niigata University,
Ikarashi, Niigata 950--2181, Japan}
\\[0.5cm]

%  \vskip1.8cm
%  {\bf Abstract:} 
\parbox[t]{\textwidth}{{\bf Abstract:}  
Cross section formulas 
for the $\Lambda$ polarization in $pp\to\Lambda^\uparrow(\ell_T)X$ and
$ep\to\Lambda^\uparrow(\ell_T)X$ are derived
and its characteristic features are discussed. 
}

\end{center}

\setcounter{equation}{0}

In this report we discuss the polarization of
$\Lambda$ hyperon produced 
in unpolarized $pp$ and $ep$ collisions relevant for the ongoing 
RHIC-SPIN, HERMES and COMPASS experiments. 
According to the QCD factorization theorem,
the polarized cross section for $pp \to \Lambda^\uparrow X$  
consists of two twist-3 contributions:
\beq
&(A)&\qquad E_a(x_1,x_2)\otimes q_b(x')\otimes \delta \widehat{q}_{c}(z)\otimes
\hat{\sigma}_{ab\to c},\nonumber\\[-5pt]
&(B)&\qquad  q_a(x)\otimes q_b(x') \otimes \widehat{G}_{c}(z_1,z_2)
\otimes 
\hat{\sigma}_{ab\to c}',\nonumber
\eeq
where 
the functions $E_a(x_1,x_2)$
and $\widehat{G}_{c}(z_1,z_2)$
are the twist-3 quantities representing, respectively, the
unpolarized distribution in the nucleon
and the polarized fragmentation function for $\Lambda^\uparrow$. 
$\delta \widehat{q}_c(x)$ is the transversity 
fragmentation function for $\Lambda^\uparrow$.
$a$, $b$ and $c$ stand for the parton's species, sum over which
is implied.  $E_a$ and $\delta\widehat{q}_c$ are chiral-odd.   
Corresponding to the above
(A) and (B), the polarized cross section for $ep \to \Lambda^\uparrow X$
(final electron is not detected) receives two twist-3 contributions:
\beq
&(A')&\qquad E_a(x_1,x_2)\otimes 
\delta\widehat{q}_{a}(z)\otimes\hat{\sigma}_{ea\to a}, \nonumber\\[-5pt]
&(B')&\qquad q_a(x)\otimes\widehat{G}_{a}(z_1,z_2)\otimes
\hat{\sigma}_{ea\to a}'.\nonumber
\eeq
The (A) contribution for $pp \to \Lambda^\uparrow X$
has been analyzed in \cite{KK01}, where
it was shown that (A) gives rise to growing $P_\Lambda$ at large $x_F$ 
as observed experimentally. 
Here we extend the study to the (B) term (see also \cite{Koike01})
at RHIC energy and also for the $ep$ collision.

The unpolarized
twist-3 distribution
$E_{F,D}(x_1,x_2)$ is defined in \cite{KK01}.
Likewise 
the twist-3 fragmentation function for 
a polarized $\Lambda$ (with momentum $\ell$) 
is defined as
the lightcone correlation function as ($w^2=0$, $\ell\cdot w =1$)
\beq
& &{1\over N_c}\sum_X\int {d\lambda\over 2\pi}\int {d\mu\over 2\pi}
e^{-i{\lambda\over z_1}}e^{-i\mu({1\over z_2}-{1\over z_1})}
\la 0 |\psi_i(0)|\pi X \ra \la \pi
X|g F^{\alpha\beta}(\mu w)w_\beta
\bar{\psi}_j(\lambda w)|0\ra\nonumber \\
& &\qquad= {M_N\over 2z_2}\,\left(\ellslash\right)_{ij}\epsilon^{
\alpha\ell w S_\perp}\widehat{G}_F(z_1,z_2)
+i{M_N\over 2z_2}\,\left(\gamma_5\ellslash\right)_{ij}S_\perp^\alpha
\widehat{G}_F^5(z_1,z_2)
+\cdots.
\label{GFhat}
\eeq
Note that we use the nucleon mass $M_N$ to normalize the twist-3
fragmentation function for $\Lambda$.
There is another twist-3 fragmentation functions
which are obtained from (\ref{GFhat}) by shifting
the gluon-field strength from the
left to the right of the cut.
The defined functions $\widehat{G}_{FR}(z_1,z_2)$ and 
$\widehat{G}_{FR}^5(z_1,z_2)$
are connected to $\widehat{G}_{F}(z_1,z_2)$ by the relation
$\widehat{G}_F(z_1,z_2) = \widehat{G}_{FR}(z_2,z_1)$ and
$\widehat{G}_F^5(z_1,z_2) = -\widehat{G}_{FR}^5(z_2,z_1)$,
which follows from hermiticity and time reversal invariance. 
Unlike the twist-3 distributions, the twist-3 fragmentation function
does not have definite symmetry property.
Another class of twist-3 fragmentation functions $\widehat{G}_D^{(5)}(z_1,z_2)$
is also defined from (\ref{GFhat}) by replacing 
$g F^{\alpha\beta}(\mu w)w_\beta$ by $D^\alpha(\mu w)=
\partial^\alpha-igA(\mu w)$.  Note, however, this is not independent
from the above (\ref{GFhat}).  

Following the method of \cite{QS99} we present the analysis of the (C) term.
The detailed analysis shows $\widehat{G}_F(z,z)$
appears as soft-gluon-pole contribution ($z_1=z_2=z$), 
while 
$\widehat{G}_D(z_1,z_2)$ appears as a soft fermion pole
($z_1=0$ or $z_2=2$).  Physically, the latter 
is expected to be suppressed, and we include only the former contribution.
This observation also applies to $E_{F,D}(x_1,x_2)$
relevant for the (A) term.
In the large $x_F$ region, 
the main contribution comes from large-$x$ and large-$z$ (and small $x'$)
region.
Since $E_F$ and $\widehat{G}_F$ behaves as $E_F(x,x)\sim (1-x)^\beta$
and $\widehat{G}_F(z,z)\sim (1-z)^{\beta'}$ with $\beta$, $\beta'>0$,
$|(d/dx)E_F(x,x)|\gg |E_F(x,x)|$,
$|(d/dz)\widehat{G}_F(z,z)| \gg |\widehat{G}_F(z,z)|$ at
large $x$ and $z$.  In particular, the valence component of
$E_F$ and $\widehat{G}_F$ dominates in this region.
We thus keep only the valence quark contribution for the derivative
of these soft-gluon pole function (``valence-quark soft-gluon
approximation'') for the $pp$ collision.  For the $ep$ case,
we include all the soft-gluon pole contribution, since the calculation
is relatively simple compared to the $pp$ case.

In general
$P_\Lambda$ is a function of 
$S=(P+P')^2\simeq 2P\cdot P'$,
$T=(P-\ell)^2\simeq -2P\cdot \ell$ and
$U=(P'-\ell)^2\simeq -2P'\cdot \ell$ where $P$ and $P'$ are the momenta
of the two nucleons, and $\ell$
is the momentum of $\Lambda$.
In the following we use
$S$, $x_F = {2\ell_{\parallel}\over \sqrt{S}} = {T-U\over S}$ and
$x_T = {2\ell_{T}\over \sqrt{S}}$ as independent variables.
The polarized cross section for the (B) term reads
\beq
& &E_\Lambda{d^3\Delta\sigma(S_\perp) \over d \ell^3}
= {2\pi M_N \alpha_s^2 \over S}\epsilon^{\alpha\ell w S_\perp}
\sum_{a}\int_{z_{min}}^1
{d\,z\over z^2}
\int_{x_{min}}^1 {d\,x\over x}
{1\over xS + U/z}
\int_0^1 {d\,x'\over x'}
\nonumber\\
& &\qquad\times \delta\left(x'+{xT/z \over xS + U/z}\right)\nonumber\\
& &\qquad\times\left\{
\sum_{b,c} q^a(x) q^b(x') 
\left[-z_1^2 {\partial \over \partial z_1}
\widehat{G}_{F}^a(z_1,z)\right]_{z_1=z}
\left({-2p_\alpha\over T}\widehat{\sigma}_{ab\to c}^I+
{-2p'_\alpha\over U}\widehat{\sigma}_{ab\to c}^{II}\right)\right.\nonumber\\
& &\qquad\left.+\sum_{b,c}  q^a(x) q^b(x') 
\left[ -z^2 {d \over d z}
\widehat{G}_{F}^a(z,z)\right]
{xp_\alpha+x'p'_\alpha \over |xT+x'U|}
\left(\widehat{\sigma}_{ab\to c}^I+
\widehat{\sigma}_{ab\to c}^{II}\right)\right.\nonumber\\
& &\qquad\left.+ q^a(x) G(x') 
\left[-z_1^2 {\partial \over \partial z_1}
\widehat{G}_{F}^a(z_1,z)\right]_{z_1=z}
\left({-2p_\alpha\over T}\widehat{\sigma}_{ag\to a}^I+
{-2p'_\alpha\over U}\widehat{\sigma}_{ag\to a}^{II}\right)\right.\nonumber\\
& &\qquad\left.+ q^a(x) G(x') 
\left[ -z^2 {d \over d z}
\widehat{G}_{F}^a(z,z)\right]
{xp_\alpha +x'p'_\alpha \over |xT+x'U|}
\left(\widehat{\sigma}_{ag\to a}^I+
\widehat{\sigma}_{ag\to a}^{II}\right)\right\},
\label{final}
\eeq
where the lower limits for the integration variables are
$z_{min} = -(T+U)/S=\sqrt{x_F^2+x_T^2}$ and 
$x_{min} = -U/z(S+T/z)$.
The partonic hard cross sections are written in terms of 
the invariants in the parton level,
$\shat =(xp + x'p')^2 =xx'S$,
$\that =(xp-{\ell/z})^2 =xT/z$ and
$\uhat =(x'p'-{\ell/z})^2 =x'U/z$.  They read
\beq
& &\widehat{\sigma}_{ab\to c}^I=-{1\over 36}{\shat^2+\uhat^2\over \that^2}
\delta_{ac} +{7\over 36}
{\shat^2+\that^2\over \uhat^2}\delta_{bc}
+{1\over 54}{\shat^2\over \that\uhat}\delta_{ab}\delta_{ac},\nonumber\\
& &\widehat{\sigma}_{ab\to c}^{II}={7\over 36}{\shat^2+\uhat^2\over \that^2}
\delta_{ac} -{1\over 36}
{\shat^2+\that^2\over \uhat^2}\delta_{bc}
+{1\over 54}{\shat^2\over \that\uhat}\delta_{ab}\delta_{ac},\nonumber\\
& &
\widehat{\sigma}_{a\bar{b}\to c}^I=-{1\over 36}{\shat^2+\uhat^2\over \that^2}
\delta_{ac} +{7\over 36}
{\uhat^2+\that^2\over \shat^2}\delta_{ab},\quad
\widehat{\sigma}_{a\bar{b}\to c}^{II}
={1\over 18}{\shat^2+\uhat^2\over \that^2}
\delta_{ac} +{1\over 18}
{\uhat^2+\that^2\over \shat^2}\delta_{ab},\nonumber\\
& &
\widehat{\sigma}_{qg\to q}^{I}={-1\over 8}\left(1-{\shat\uhat\over \that^2}
\right)+
{1\over 288}\left({-\uhat\over \shat}+{\shat\over -\uhat}\right)
-{\shat\over 16\that}-{\uhat\over 16\that},\nonumber\\
& &
\widehat{\sigma}_{qg\to q}^{II}={9\over 16}\left(1-{\shat\uhat\over \that^2}
\right)+
{\uhat\over 32\shat}-{\shat\over 4\uhat}+{9\uhat\over 16\that}.
\label{parton}
\eeq
Among these partonic cross sections, $\widehat{\sigma}^{I}$ becomes 
more important
at large $x_F$ because of the $1/T$ factor in (\ref{final}).

To estimate the above contribution,
we introduce a model ansatz as
$\widehat{G}_F^a(z,z)=K_a \widehat{q}_{a}(z)$ with twist-2 unpolarized 
fragmentation function $\widehat{q}^a(z)$, noting that
the Dirac structure of $\widehat{G}_F^a(z,z)$ and $\widehat{q}_{a}(z)$ 
is the same\,\cite{QS99}.  
$K_a$'s are taken to be $K_u=-K_d=0.07$ which are the same
values used in the relation $G_F(x,x)=K_aq^a(x)$ to
reproduce $A_N$ in $p^\uparrow p\to \pi X$ observed at E704\,\cite{KK00}.
As noted before, 
$\widehat{G}_F(z_1,z_2)$ does not have definite symmetry property
unlike the twist-3 distribution $E_F(x_1,x_2)$. 
Nevertheless we assume 
$\left[(\partial/\partial z_1)\widehat{E}_F(z_1,z)\right]_{z_1=z}
=(1/2)(d/d z)\widehat{E}_F(z,z)$.  
The result for the $\Lambda$ polarization $P_\Lambda^{pp}$ 
at $\sqrt{S}=62$ GeV
is shown in 
Fig. 1 together with the R608 data.
There (A) (chiral-odd) contribution 
studied in \cite{KK01} is also shown for comparison.
(For the adopted distribution and fragmentation
functions, see \cite{KK01}.)  One sees that the tendency 
of $P_\Lambda^{pp}$ from
the (B)(chiral-even) contribution is quite similar to the R608 data.
Rising behavior of $P_\Lambda^{pp}$ at large $x_F$ comes from
(i) the large partonic cross sections in 
(\ref{parton}) ($\sim 1/\hat{t}^2$ term)
and (ii) the derivative of the soft-gluon pole functions.  
With these parameters $K_a$,
$P_\Lambda^{pp}$ at RHIC energy ($\sqrt{S}=200$ GeV)
is shown in Fig.2 at $l_T=1.5$ GeV.  
Fig. 3 shows the $l_T$ dependence of $P_\Lambda^{pp}$ of the (B) term,
indicating large $\ell_T$ dependence at $1\leq \ell_T \leq 3$ GeV.   
Experimentally, $P_\Lambda^{pp}$ grows up as $\ell_T$ increases up to
$\ell_T \sim 1$ GeV and stays constant at $1\leq \ell_T\leq 3$ GeV.
So the $P_\Lambda^{pp}$ observed at R608 can not be wholly ascribed to
the twist-3 effect studied here which is designed to
describe large $\ell_T$ polarization. 
\begin{figure}[htb]
\setlength{\unitlength}{1cm}
\begin{minipage}[t]{7.0cm}
\begin{picture}(6.5,6.5)
%\hspace*{-0.2cm}
\psfig{file=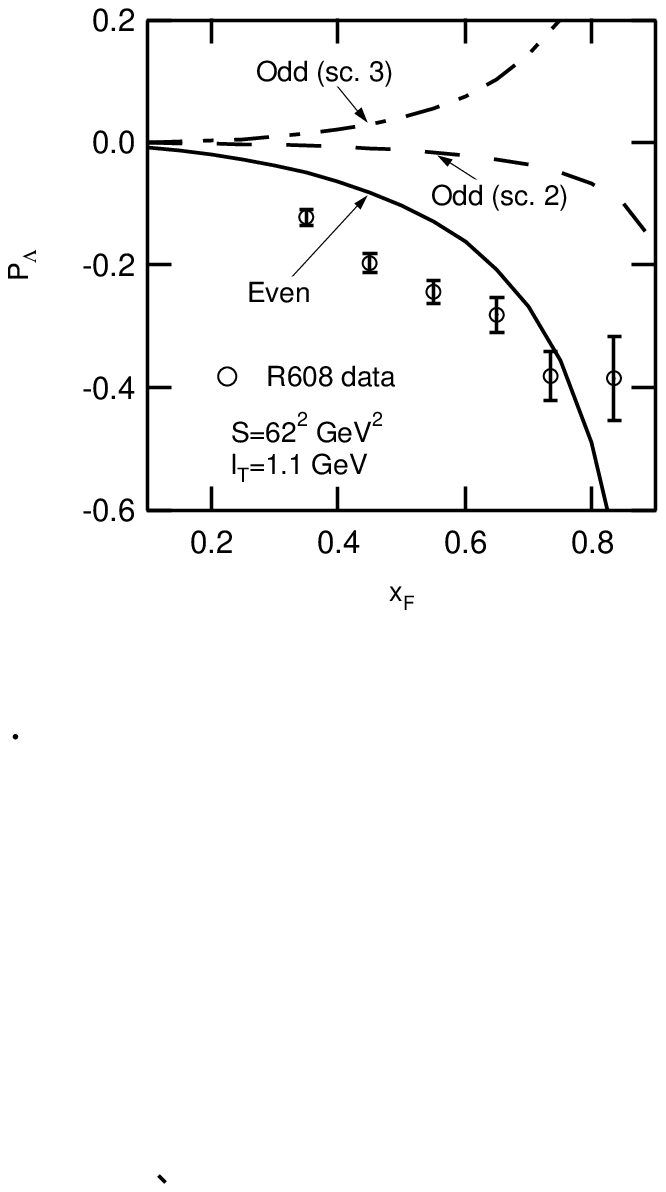,width=6.6cm,height=7cm}
\end{picture}\par
\vspace{-1.6cm}
\caption{$P_\Lambda^{pp}$ at $\sqrt{S}=62$ GeV.}
\label{fig:1}
\end{minipage}
\hfill
\begin{minipage}[t]{7.0cm}
\begin{picture}(6.5,6.5)(0,-1.1)
%\hspace*{-0.2cm}
\psfig{file=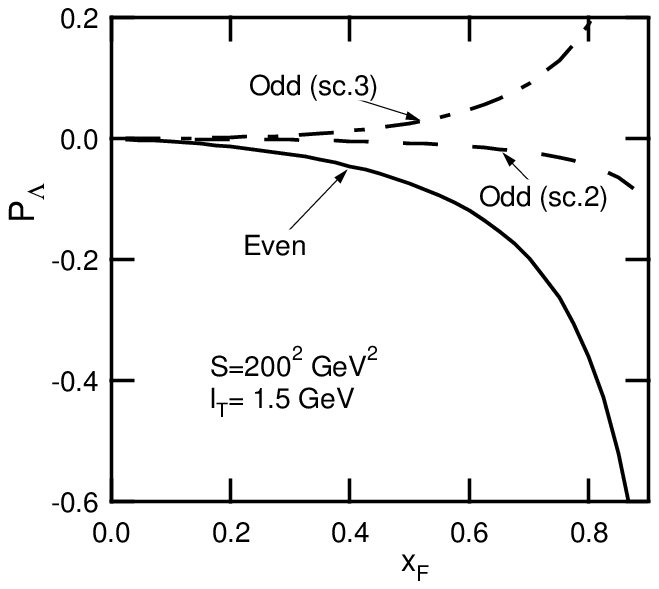,width=6.1cm,height=5.8cm}
\end{picture}\par
\vspace{-1.7cm}
\caption{$P_\Lambda^{pp}$ at $\sqrt{S}=200$ GeV.}
\label{fig:2}
\end{minipage}
\end{figure}

\vspace{-1cm}
We next discuss the polarization $P_\Lambda^{ep}$ in
$p e\to \Lambda^\uparrow(\ell)X$
where the final electron is not observed.  
In our $O(\alpha_s^0)$ calculation, 
the exchanged photon remains highly virtual
as far as the observed $\Lambda$ has
a large transverse momentum $\ell_T$ with respect to the $ep$ axis.
Therefore experimentally one needs to integrates only over those 
virtual photon events to compare with our formula.
\begin{figure}[htb]
\setlength{\unitlength}{1cm}
\begin{minipage}[t]{7.0cm}
\begin{picture}(6.5,6.5)(0,-1.1)
%\hspace*{-0.2cm}
\psfig{file=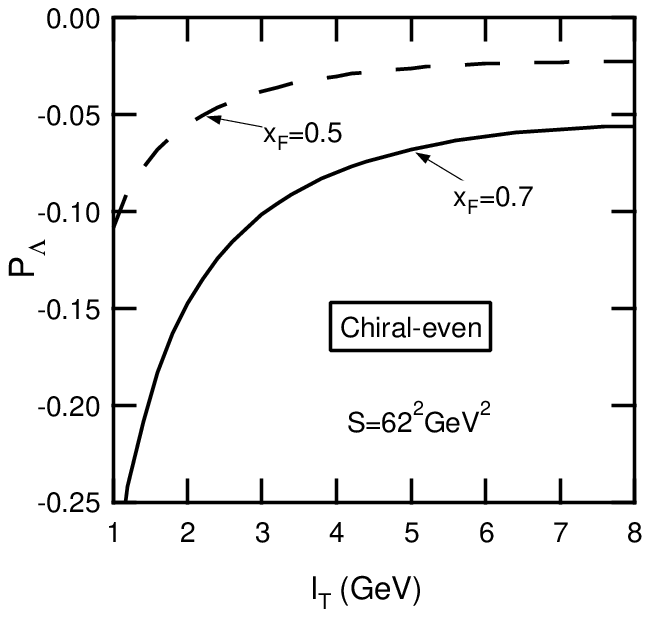,width=6.5cm,height=5.8cm}
\end{picture}\par
\vspace{-1.7cm}
\caption{$\ell_T$ dependence of $P_\Lambda^{pp}$.}
\label{fig:3}
\end{minipage}
\hfill
\begin{minipage}[t]{7.0cm}
\begin{picture}(6.5,6.5)(0,-1.1)
%\hspace*{-0.2cm}
\psfig{file=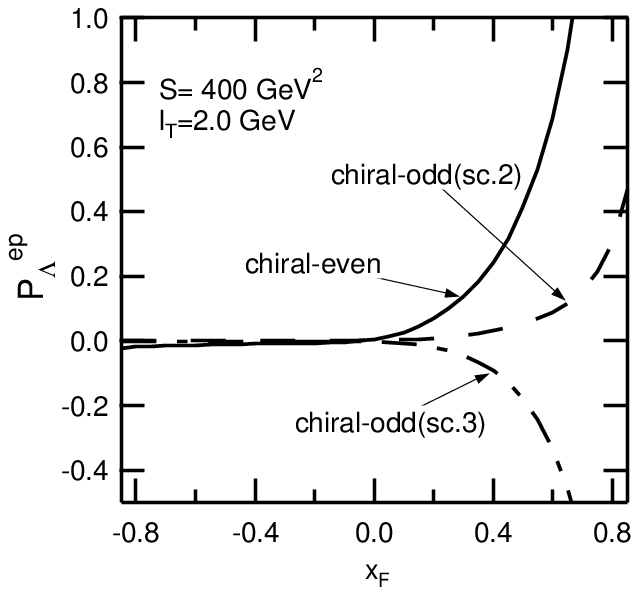,width=6.5cm,height=5.8cm}
\end{picture}\par
\vspace{-1.7cm}
\caption{$P_\Lambda^{ep}$ at $\sqrt{S}=20$ GeV.}
\label{fig:4}
\end{minipage}
\end{figure}

\vspace{-1cm}
Using the twist-3 distribution and fragmentation functions used
to describe $P_\Lambda^{pp}$,
we show in Fig. 4 the obtained 
$P_\Lambda^{ep}$ corresponding to
(A')(chiral-odd) and (B')(chiral-even) contributions.
Remarkable feature of Fig. 4 is that in both chiral-even and
chiral-odd contributions (i) the sign of $P_\Lambda^{ep}$
is opposite to the sign of $P_\Lambda^{pp}$ and (ii) the magnitude of 
$P_\Lambda^{ep}$
is much larger than that of $P_\Lambda^{pp}$, in particular, at large $x_F$,
and it even overshoots one.
(In our convention, $x_F >0$ 
corresponds to the production of $\Lambda$ in the forward hemisphere of
the initial proton in the $ep$ case.)
The origin of these features can be traced back to the color factor in
the dominant diagrams for the {\it twist-3 polarized} cross sections
in $ep$ and $pp$ collisions.
Of course, the $P_\Lambda$ can not exceeds one, and thus our model 
estimate needs to be modified.  
First, the applied kinematic range
of our formula should be reconsidered: 
Application of the twist-3 cross section
at such small $\ell_T$ may not be justified.
Second, our simple model ansatz of $E_F^a(x,x)\sim \delta q^a(x)$
(in (A) term) and 
$\widehat{G}_F^a(z,z)\sim \widehat{q}^a(z)$ 
should be modified at $x\to 1$ and
$z\to 1$, respectively.  The derivative of these functions,
which is important for the growing $P_\Lambda^{pp}$ at large $x_F$, 
eventually
leads to divergence of $P_\Lambda$ at $x_F\to 1$ as $\sim 1/(1-x_F)$.

As a possible remedy for this 
pathology we tried the following:  As an example for the (B) (chiral-even)
contribution
we have a model $\widehat{G}_F^a(z,z)\sim \widehat{q}_a(z)\sim_{z\to 1} 
(1-z)^{\beta}$
where $\beta=1.83$ in the fragmentation function we adopted.
Tentatively we shifted $\beta$ as $\beta\to\beta(z)=\beta+z^8$,
which suppresses the divergence of $P_\Lambda$ at $x_F\to 1$ but still
keeps rising behavior of $P_\Lambda$ at large $x_F$.  This avoids 
overshooting of one in $P_\Lambda^{ep}$ 
but reduces $P_\Lambda^{pp}$ seriously.
The result obtained by this modification is shown in Figs. 5 and 6.

To summarize we have studied the $\Lambda$ polarization
in $pp$ and $ep$ collisions in the framework of collinear factorization.
Our approach includes all effects for the large $\ell_T$ production.
One needs to be causious in
interpreting the available $pp$ data at relatively low $\ell_T$
in terms of the derived formula. 
Determination of the participating twist-3 functions requires
global analysis of future $pp$ and $ep$ data.

\begin{figure}[htb]
\setlength{\unitlength}{1cm}
\begin{minipage}[t]{7.0cm}
\begin{picture}(6.5,6.5)
\hspace*{-0.2cm}
\psfig{file=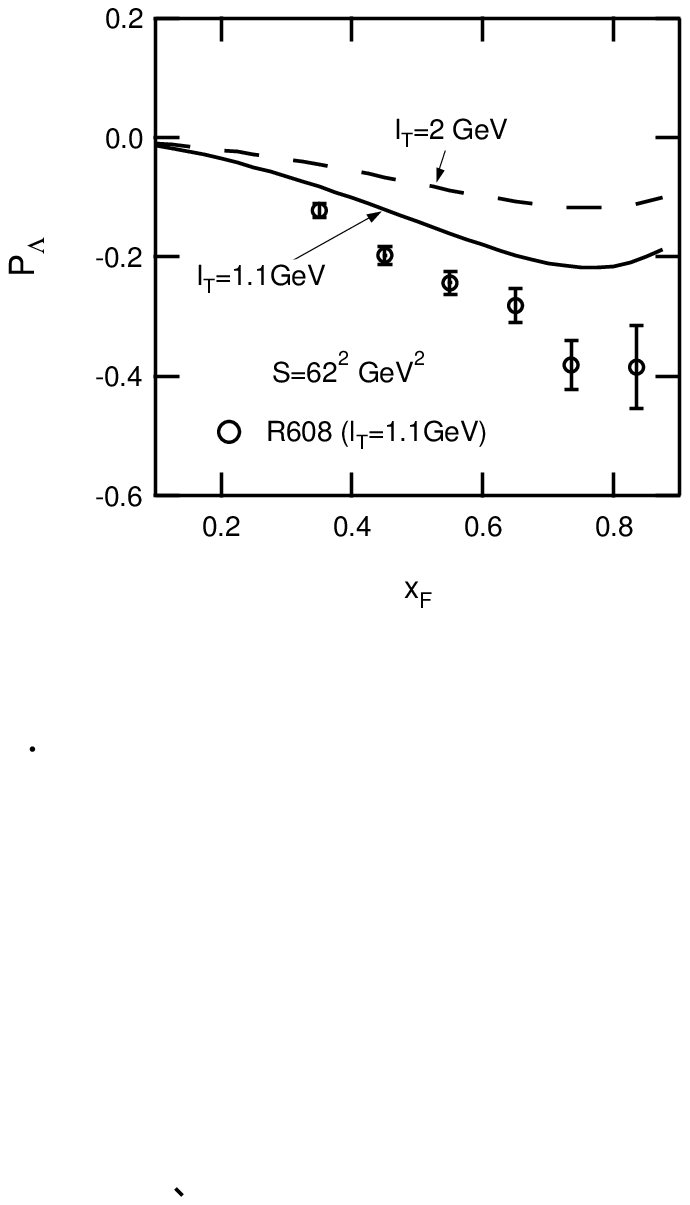,width=6.5cm,height=7cm}
\end{picture}\par
\vspace{-1.7cm}
\caption{$P_\Lambda^{pp}$ with modified $\widehat{G}_F$.}
\label{fig:5}
\end{minipage}
\hfill
\begin{minipage}[t]{7.0cm}
\begin{picture}(6.5,6.5)(0,-1.1)
\hspace*{-0.2cm}
\psfig{file=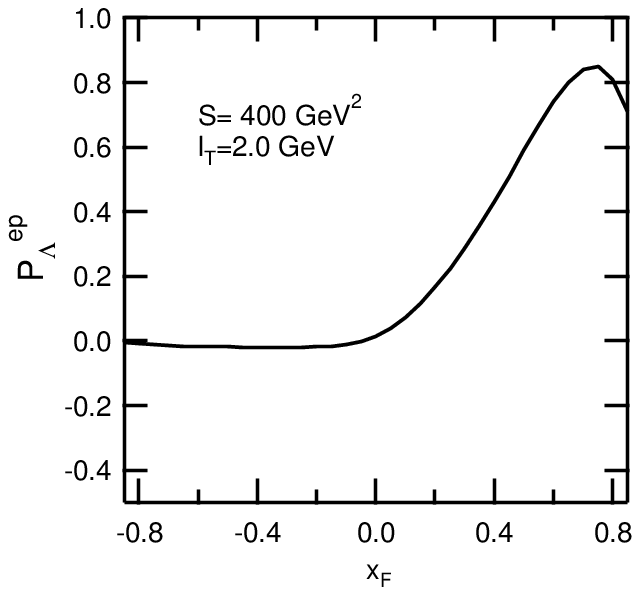,width=6.5cm,height=5.8cm}
\end{picture}\par
\vspace{-1.7cm}
\caption{$P_\Lambda^{ep}$ with modified $\widehat{G}_F$.}
\label{fig:6}
\end{minipage}
\end{figure}

\vspace{-0.8cm}

\noindent
{\bf Acknowledgement:} This work is supported in part by the 
Grant-in-Aid for Scientific Research of Monbusho.

\end{document}